\documentclass[reprint, aps, prx, floatfix, amsmath, amssymb]{revtex4-1}
\usepackage{float, graphicx, hyperref}
\usepackage{bm, bbm, physics, braket}
\hypersetup{colorlinks=true,urlcolor=blue,linkcolor=blue,citecolor=blue,}

\newcommand{\bk}{{\bm k}}
\newcommand{\br}{{\bm r}}
\newcommand{\bd}{{\bm d}}
\newcommand{\mP}{{\mathcal P}}
\newcommand{\mHI}{\mathcal{H}_\text{I}}
\newcommand{\mHMF}{\mathcal{H}_{\text{MF}}}
\newcommand{\mH}{\mathcal{H}_{0}}
\newcommand{\psiMF}{\Psi_{\text{MF}}}
\newcommand{\gMF}{g_{c}^{\text{MF}}}

\begin{document}

\title{Triangular lattice Majorana-Hubbard model: \\ Mean field theory and DMRG on a width-4 torus}

  \author{Tarun Tummuru}
  \affiliation{Department of Physics and Astronomy \& Stewart Blusson Quantum Matter Institute,
  University of British Columbia, Vancouver, British Columbia, Canada, V6T 1Z4}

  \author{Alberto Nocera}
  \affiliation{Department of Physics and Astronomy \& Stewart Blusson Quantum Matter Institute,
  University of British Columbia, Vancouver, British Columbia, Canada, V6T 1Z4}

  \author{Ian Affleck}
  \affiliation{Department of Physics and Astronomy \& Stewart Blusson Quantum Matter Institute,
  University of British Columbia, Vancouver, British Columbia, Canada, V6T 1Z4}

  \date{\today}


\begin{abstract}
  Majorana modes can arise as zero energy bound states in a variety of solid state systems. A two-dimensional phase supporting these quasiparticles, for instance, emerges on the surface of a topological superconductor with the zero modes localized at the cores of vortices. At low energies, such a setup can be modeled by Majorana modes that interact with each other on the Abrikosov lattice. In experiments, the lattice is usually triangular. Motivated by the practical relevance, we explore the phase diagram of this Hubbard-like Majorana model using a combination of mean field theory and numerical simulation of thin torus geometries through the density matrix renormalization group algorithm. Our analysis indicates that attractive interactions between Majoranas can drive a phase transition in an otherwise gapped topological state.
\end{abstract}

\maketitle


\section{Introduction}

  The past decade has seen tremendous progress in the quest for realizing a localized Majorana fermion in the laboratory. The so-called Majorana zero mode (MZM) is a zero energy mid-gap excitation that arises as a localized quasiparticle in some low dimensional systems \cite{kitaev2001unpaired, Oreg_2010, Sau_2010, Lutchyn_2010, fu2008superconducting}. A defect supporting a MZM is perhaps the simplest manifestation of a non-Abelian anyon. And since a non-Abelian anyon is the key ingredient in braiding based topological quantum computation, Majoranas are a subject of topical interest \cite{Nayak_2008, Elliott_2015}.

  Vortices in a 2D superconductor with chiral $p$-wave pairing harbor MZMs at their cores \cite{ivanov2001non}. Fu and Kane showed that such an effective $p$-wave pairing can be realized at the interface of a $s$-wave superconductor and a strong topological insulator \cite{fu2008superconducting}. Zero modes that are completely isolated from each other are ideal from the perspective of implementing unitary quantum gates via braiding vortices. In practice, though, the MZMs are exponentially localized at best, with a scale set by the coherence length of the superconductor under question, thereby causing the MZM wavefunctions to overlap \cite{cheng2010tunneling}. When the mid-gap states are well separated from the rest of the quasiparticle spectrum, at thermal energy scales below the gap, the effective Hamiltonian describing the zero modes is a sum of local terms involving pairwise Majorana operators. In the presence of a vortex lattice with a finite density of zero modes, it is then natural to describe the system with a Hubbard-like tight-binding model for MZMs.

  One appealing feature of the Fu-Kane proposal is that the topological insulator's chemical potential $\mu$ can be tuned to control the MZM wavefunction overlaps. In particular, when $\mu$ coincides with the surface state's Dirac point, the interface superconductor exhibits an emergent chiral symmetry that prevents the Majorana modes from hybridizing \cite{teo2010topological, chiu2015strongly}. In the vicinity of this neutrality point, with the single particle tunneling amplitudes greatly reduced, four fermion terms are the leading perturbation and the system is, therefore, strongly interacting.

  The prospect of realizing interacting Majorana models has opened doors to a host of exotic proposals and predictions \cite{Nussinov_2012, rahmani2019interacting}. Extensive studies of 1D MZM chains have found that these models show supersymmetry, with some exhibiting phase transitions belonging to the tri-critical Ising universality class \cite{Grover_2014, rahmani2015emergent, hsieh2016all, sannomiya2019supersymmetry, o2018lattice, Li_2020}. In 2D, Majorana Hamiltonians on square \cite{affleck2017majorana, wamer2018renormalization, kamiya2018majorana}, kagome \cite{li2019supersymmetry} and honeycomb lattices \cite{li2018majorana} also have interesting phase diagrams. Further, lattices of MZMs have been shown to enable new schemes of surface code quantum computation \cite{bravyi2010majorana, vijay2015majorana}.

  On the experimental front, there is now promising evidence for zero modes in the vortex cores of topological superconductors \cite{Xu_2015, wang2018evidence, Liu_2018, machida2019zero}. A common feature of such experiments is that the Abrikosov lattice is triangular; as the best packed lattice in 2D, this arrangement accommodates the maximum inter-vortex separation. Given this context, a study of MZMs on the triangular lattice is highly relevant. A few earlier works have considered the role of disorder in this setup at a non-interacting level \cite{kraus2011majorana, laumann2012disorder, lahtinen2014perturbed, chiu2020scalable}. Our objective here is to understand the role of interactions. Pursuant to this goal, we analyze plausible spontaneous symmetry breaking within the framework of a self-consistent mean field theory (Sec.\,\ref{sec:SC_MF}) and study the model numerically using the density matrix renormalization group (DMRG) algorithm \cite{white1992density, white1993density, schollwock2011density} (Sec.\,\ref{sec:4leg_lad}). To begin, we introduce the model and discuss its symmetries.


\section{The model} \label{chap:mod_and_symm}

\begin{figure}
    \centering
    \includegraphics[width=8cm]{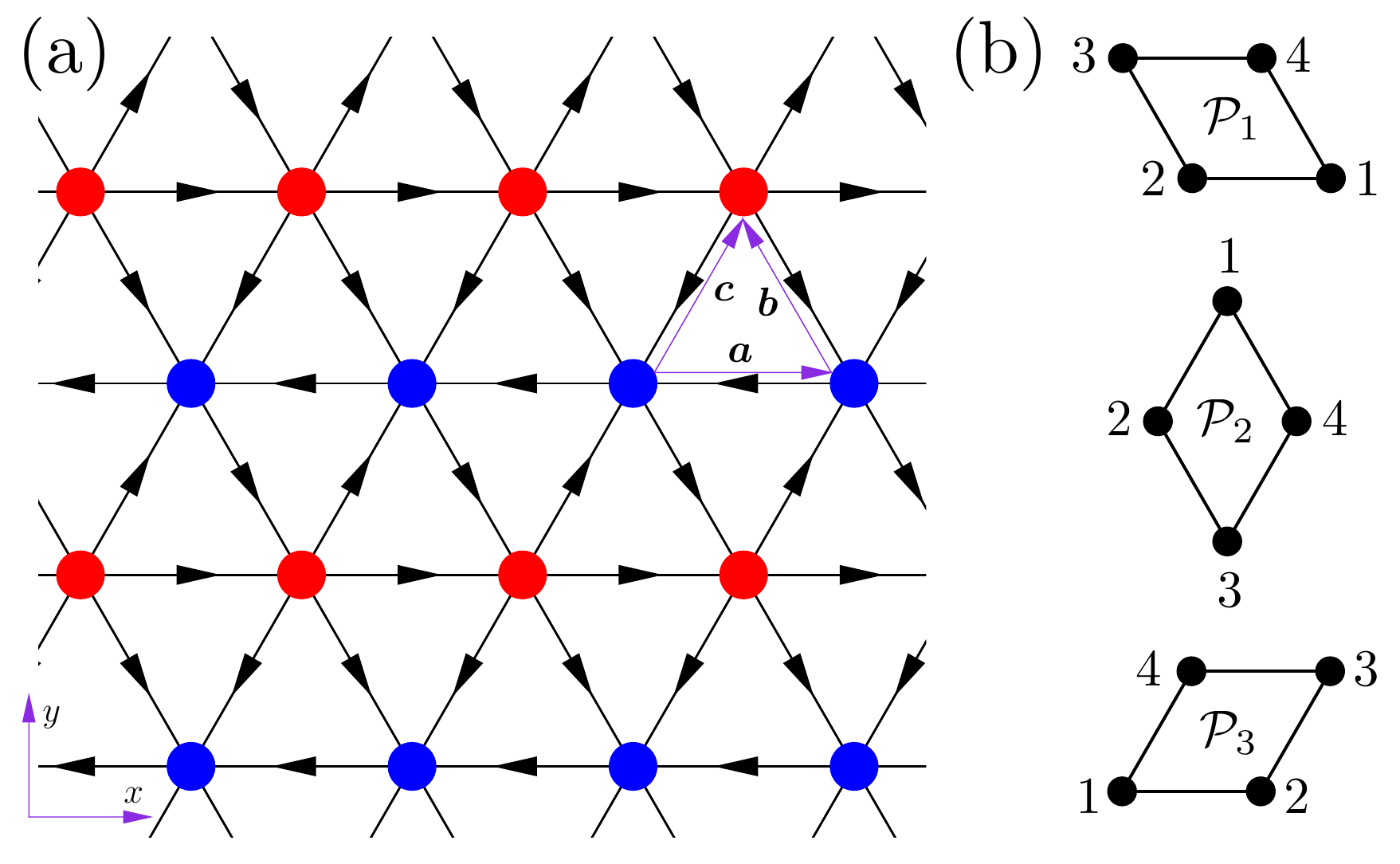}
    \caption{(a) A $\mathbb{Z}_2$ gauge for the triangular lattice that we adopt in this work. Hopping along (against) the directions indicated incurs a phase of $+i$ $(-i)$. The two sub-lattices of the rectangular Bravais lattice are shown in red and blue. (b) Ordering of MZM operators in the three plaquette interaction terms.}
    \label{fig:lattice1}
  \end{figure}

  The self-adjoint nature of Majorana operators $\gamma_p = \gamma^\dagger_p$, together with the requirement of having a Hermitian Hamiltonian, dictates that a lattice hosting MZMs is described by
  \begin{align}
    \mH = it \sum_{\langle pq \rangle} \eta_{pq} \gamma_p \gamma_q,
    \label{eq:H0}
  \end{align}
  where $p$ and $q$ label nearest neighbor sites and the fermionic operators obey $\{\gamma_p, \gamma_q\} = 2 \delta_{pq}$. The purely real prefactor $t$ is interpreted as the probability amplitude for a quasiparticle to tunnel between two vortices. The anti-symmetric matrix $\eta_{pq} = \pm 1$ indicates the sign of the phase $i$ acquired in a tunneling process. At this point, the choice of $\eta_{pq}$ is arbitrary because one may redefine $\gamma_p \to -\gamma_p$ without altering the MZM anti-commutation relations. The ambiguity can be understood as a $\mathbb{Z}_2$ gauge freedom inherent to the system. The product of phases along a closed loop, however, corresponds to $\mathbb{Z}_2$ flux and is gauge invariant -- a fact that is encapsulated in the Grosfeld-Stern rule \cite{grosfeld2006electronic}. Fig.\,\ref{fig:lattice1}(a) shows one possible gauge choice that is relevant to Majoranas on a triangular vortex lattice \cite{kraus2011majorana, liu2015electronic}; the corresponding $\mH$ has been written out explicitly in Appendix \ref{sec:H_cf}. Note that gauge fixing imposes a rectangular Bravais lattice with a two-site unit cell. An alternate gauge would modify $\eta_{pq}$, but would not reduce the number of the sub-lattice degrees of freedom. With this Hamiltonian as the starting point, we shall use its symmetries to determine the form of interactions.

  \subsection{Symmetries}

    Due to the $\mathbb{Z}_2$ gauge, $\mH$ is not always manifestly invariant under lattice transformations. For example, translation by one site along the direction $\bm{c}$ (see inset of Fig.\,\ref{fig:lattice1}(a) for the directions referred to in the following) does not map the model onto itself. The reason being that symmetries involving Majorana modes are represented projectively. Correspondingly, conventional symmetry operations should be supplemented with gauge transformations.

    While the Bravais lattice is rectangular, the symmetries of Hamiltonian \eqref{eq:H0} are dictated by the underlying triangular lattice. In addition to discrete translations $\mathcal{T}_\mu$ along the directions $\mu = \bm{a}, \bm{b}, \bm{c}$ \footnote{The three translations are not independent because any one of them can be generated using a combination of the other two.}, a $\pi/3$ rotation about any lattice site also leaves $\mH$ invariant. Though the anti-unitary time reversal operation $\Theta$ ($i \to -i$) and reflections $\mathcal{R}_{x/y}$ about $x/y$ Cartesian axes are \emph{not} symmetries by themselves, the product $\Theta \mathcal{R}_{x/y}$ commutes with the Hamiltonian. The gauge factors accompanying each of these symmetries are outlined explicitly in Appendix \ref{sec:app_symm}.


  \subsection{Interactions}

    Because the Majorana operator at any site squares to identity, interactions necessarily involve four neighboring sites. In a square lattice, for example, these are the zero modes at the corners of an elementary square \cite{affleck2017majorana}. In a triangular geometry, three different orientations of rhomboidal plaquettes are possible, with each kind tessellating the entire lattice exactly once. Summing over all such terms, we have
    \begin{align}
      \mHI = g \sum [\mP_1 + \mP_2 + \mP_3] \quad
      \text{with} \quad \mP_\nu = \gamma_p \gamma_q \gamma_r \gamma_s,
    \end{align}
    where $g$ is the interaction strength. The three kinds of plaquettes $\mP_\nu$ ($\nu=1,2,3$) and the ordering of MZM operators in each term is shown in Fig.\,\ref{fig:lattice1}(b). With this choice, it can be verified that $\mHI$ obeys all the symmetries of $\mH$. Under the action of $\Theta \mathcal{R}_x$, for instance, $\mP_1 \leftrightarrow \mP_3$ and $\mP_2$ remains invariant.

    Henceforth, we denote the full Hamiltonian as $\mathcal{H} = \mH + \mHI$ and set $t=1$, unless specified otherwise. The number of unit cells along the two independent axes will be identified by $N_x$ and $N_y$.


\section{Two limits}

  \subsection{Strong coupling}

    As mentioned previously, in a Fu-Kane realization of the MZM lattice, the topological insulator's chemical potential provides a knob to tune the zero mode overlap amplitudes. At neutrality, $t=0$ and hence $\mathcal{H} = \mHI$. In this limit, the model possesses a few interesting features that we briefly comment on.

    Because terms quadratic in the Majorana operators are absent, the $\mathbb{Z}_2$ gauge is no longer relevant and the size of the unit cell reduces to one. Since we now have an odd number of Majoranas per unit cell, periodic boundary conditions and translation symmetry dictate that the ground state is at least two-fold degenerate \cite{hsieh2016all}. It is important to note that this degeneracy is intrinsically dependent on the system's linear dimensions. For periodic systems with one odd length (either $N_x$ or $N_y$ is odd) \footnote{When the number of Majoranas per unit cell is odd, both $N_x$ and $N_y$ cannot simultaneously be odd; the Hilbert space of an odd number of Majoranas is ill-defined.}, the two degenerate states belong to different fermionic parity sectors and the degeneracy can be attributed to underlying supersymmetry. With two even lengths (both $N_x$ and $N_y$ are even), on the other hand, the degeneracy is a result of anti-commutation of translation operators along the two axes.

    When the system is defined on a torus and the number of unit cells in each direction is even, observe that changing the sign of zero mode operators at every alternate red site in Fig.\,\ref{fig:lattice1}(a) results in $\mHI$ picking up an overall negative sign, while still preserving the fermionic anti-commutation relations. Therefore, attractive and repulsive interactions are equivalent. When $N_x$ or $N_y$ is odd, however, one cannot ensure that the sign on every other red site is flipped because of periodic boundary conditions. Energy spectra, obtained numerically for small systems, confirm this reasoning.

    We emphasize that these analytical arguments do not apply when the strong coupling limit is perturbed because even an infinitesimal $t$ introduces the $\mathbb{Z}_2$ gauge.


  \subsection{Non-interacting limit}

    \begin{figure}
      \centering
      \includegraphics[width=7cm]{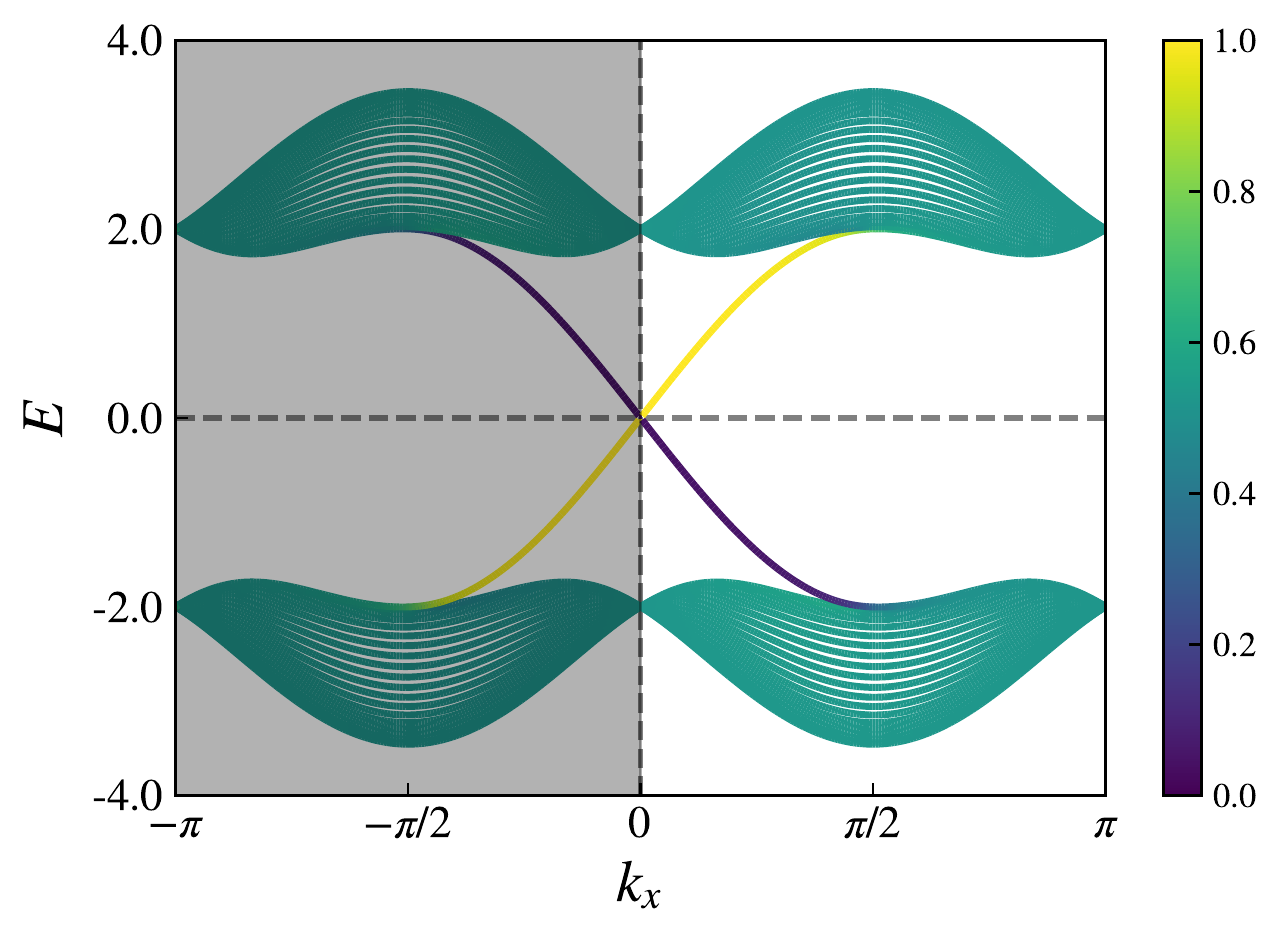}
      \caption{The energy spectrum of an infinite strip geometry with an open boundary along the $y$ direction. Non-trivial topology of the bands leads to chiral edge states that traverse the gap. The color scale denotes the normalized expectation value of the $\hat{y}$ position operator. Physically relevant momenta are marked by the unshaded region.}
      \label{fig:edge_states}
    \end{figure}

    Let us now look at the ground state properties of the model in the opposite limit, i.e., in the absence of interactions. Employing translation symmetry, it is convenient to work with momentum space operators that can be shown to obey the relation $\{\gamma_{\bk}, \gamma_{\bk'} \} = \delta_{\bk, -\bk'}$ or $\gamma_{-\bk} = \gamma^\dagger_{\bk}$. This property, which is a manifestation of the self-adjoint nature of the zero modes in real space, implies that operators at $\bk$ and $-\bk$ are not independent. Taking this into account, the Hamiltonian $\mH = \sum_{\bk}' \Psi^{\dagger}_{\bk} h_k \Psi_{\bk}$ is obtained by considering only one-half of the Brillouin zone. We indicate this with a prime over the sum. Therein, $\Psi_{\bk} = (\gamma^{r}_{\bk}, \gamma^{b}_{\bk})^{\text{T}}$ and $(r,b)$ label the two sub-lattices according to color. The Bloch Hamiltonian reads
    \begin{align}
      h_{\bk} = 2t
      \begin{pmatrix}
          -2 \sin(\bk \cdot \bd_1) & D(\bk) \\[6pt]
          D(\bk)^* & 2 \sin(\bk \cdot\bd_1)
      \end{pmatrix},
      \label{eq:bloch_h}
    \end{align}
    where $D(\bk) = i[1 - e^{-i \bk \cdot \bd_1} + e^{-i \bk \cdot \bd_2}+ e^{-i \bk \cdot (\bd_1 + \bd_2)}]$ and $\bd_1 = (1, 0)$ and $\bd_2 = (0, \sqrt{3})$ are the Bravais lattice vectors with the inter-vortex distance set to unity. Diagonalizing $h_{\bk}$ leads to the gapped dispersion
    \begin{align}
      E_{\bk}^{\pm} = \pm 2\sqrt{2} t \sqrt{3 - \cos(2 k_x) - 2 \sin(k_x) \sin(\sqrt{3} k_y)}.
      \label{eq:spec}
    \end{align}


    As a consequence of the background $\mathbb{Z}_2$ flux, a gapped band structure of MZMs can be topologically non-trivial with a nonzero Chern number $\mathcal{C}$. Indeed, we find that $\mathcal{C} = \text{sign}(t)$ \footnote{Although only half of the Brillouin zone is physically relevant, a Chern number calculation is only defined on the full periodic manifold.} and in a geometry with open boundaries the model exhibits edge states that connect the bulk bands, as depicted in Fig.\,\ref{fig:edge_states}.


    In order to understand the effect of interactions on this spectrum, in the following section we treat the weak coupling regime using mean field approximation.


\section{Mean field theory} \label{sec:SC_MF}

  Phase diagrams of previously studied Majorana-Hubbard models suggest that interactions can give rise to a Peierls-like instability and the zero modes prefer a dimerized configuration over a translationally invariant state \cite{rahmani2015phase, affleck2017majorana}. Such a hybridization between two Majoranas results in a spinless Dirac fermion state, which can be filled or empty depending on whether the interactions are attractive or repulsive. In the square lattice, for instance, translation along either $x$ or $y$ axis can be broken, thereby leading to a four-fold degenerate ground state \cite{affleck2017majorana}.

  \begin{figure}
    \centering
    \includegraphics[width=7cm]{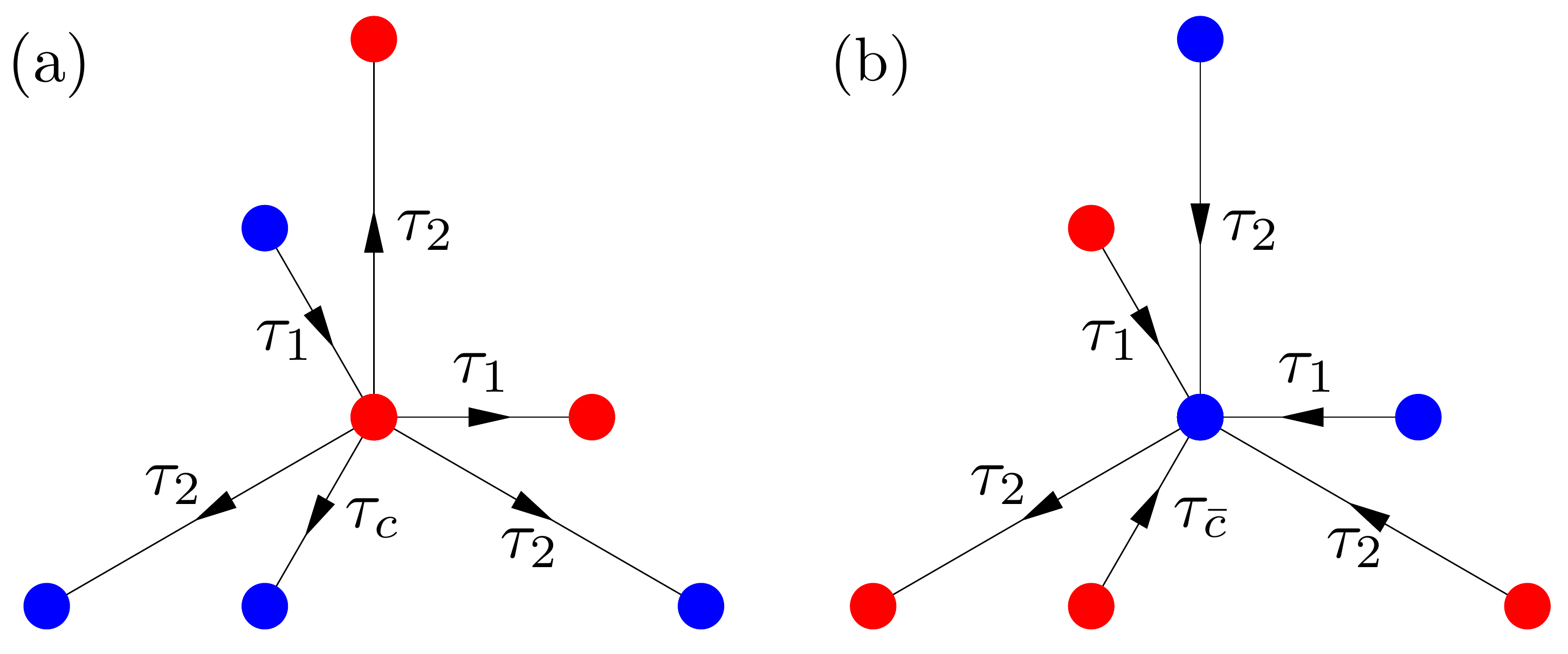}
    \caption{The mean field hopping amplitudes that connect (a) red and (b) blue sub-lattices (located at the center of each figure) to their respective nearest and next nearest neighbors. $\{\tau_{1}, \tau_c, \tau_{\bar{c}}\}$ and $\tau_2$ denote the first and second neighbor hopping amplitudes respectively.}
    \label{fig:lattice2}
  \end{figure}

  Along similar lines, the triangular lattice presents three equivalent directions $\bm{a}$, $\bm{b}$ and $\bm{c}$ for translation symmetry breaking. To explore such a tendency, we focus on the scenario where the zero modes dimerize along $\bm{c}$ \footnote{One can probe symmetry breaking along $\bm{a}, \bm{b}$ and $\bm{c}$ simultaneously by enlarging the unit cell to four sites and treating the directions on an equal footing. This is a straightforward extension of the theory presented here and it results in the same conclusions.}. In the dimerized state, one may anticipate that the tunneling amplitudes on consecutive bonds along $\bm{c}$ would differ in magnitude. We denote them by $\tau_{c}$ and $\tau_{\bar{c}}$. The rest of the first neighbor amplitudes would remain identical ($\tau_{1}$). These parameters have an intuitive origin in the mean field context: turning on interactions renormalizes the nearest neighbor hoppings from their bare value $t$. Further, a Wick's expansion of the four fermion plaquette terms shows that interactions also generate second neighbor tunneling amplitudes ($\tau_2$), which conform to the symmetries of $\mH$ and are, therefore, allowed. The parameters $\tau_j$ with $j \in \{c, \bar{c}, 1, 2\}$, shown in Fig.\,\ref{fig:lattice2}, motivate the definition of a mean field Hamiltonian
  \begin{align}
    \mHMF = i \sum_{\substack{k = c, \bar{c}, 1 \\ \langle pq \rangle}}
            \tau_k \eta_{pq} \gamma_p \gamma_q
            +
            i \tau_2 \sum_{\langle \langle pq \rangle \rangle} \eta_{pq} \gamma_p \gamma_q,
    \label{eq:HMF}
  \end{align}
  where $\gamma_p$ and $\gamma_q$ are MZM operators on the bond labeled by $\tau_j$. With the ground state wavefunction $\ket{\psiMF}$ of $\mHMF$ as a variational ansatz, minimization of the energy $\bra{\psiMF} \mathcal{H} \ket{\psiMF}$ with respect to $\tau_j$ leads to the mean field self-consistency equations (see Appendix \ref{sec:app_mft})
  \begin{align}
    \tau_c = \tau_{\bar{c}} &= t - g (2\Delta_c + 2\Delta_{\bar{c}} - \Delta_2) \nonumber \\
    \tau_1 &= t - g (4 \Delta_1 - \Delta_2) \nonumber \\
    \tau_2 &= \frac{g}{6} (4 \Delta_1 + \Delta_c + \Delta_{\bar{c}}).
    \label{eq:taus}
  \end{align}
  Therein, we have defined the expectation value of pairwise Majoranas on the bond corresponding to $\tau_j$ by $\Delta_j = \langle i \gamma_p \gamma_q \rangle$, with the convention that the operators are ordered in accordance with the direction of the $\mathbb{Z}_2$ gauge on the bond.

  \begin{figure}
    \centering
    \includegraphics[width=8cm]{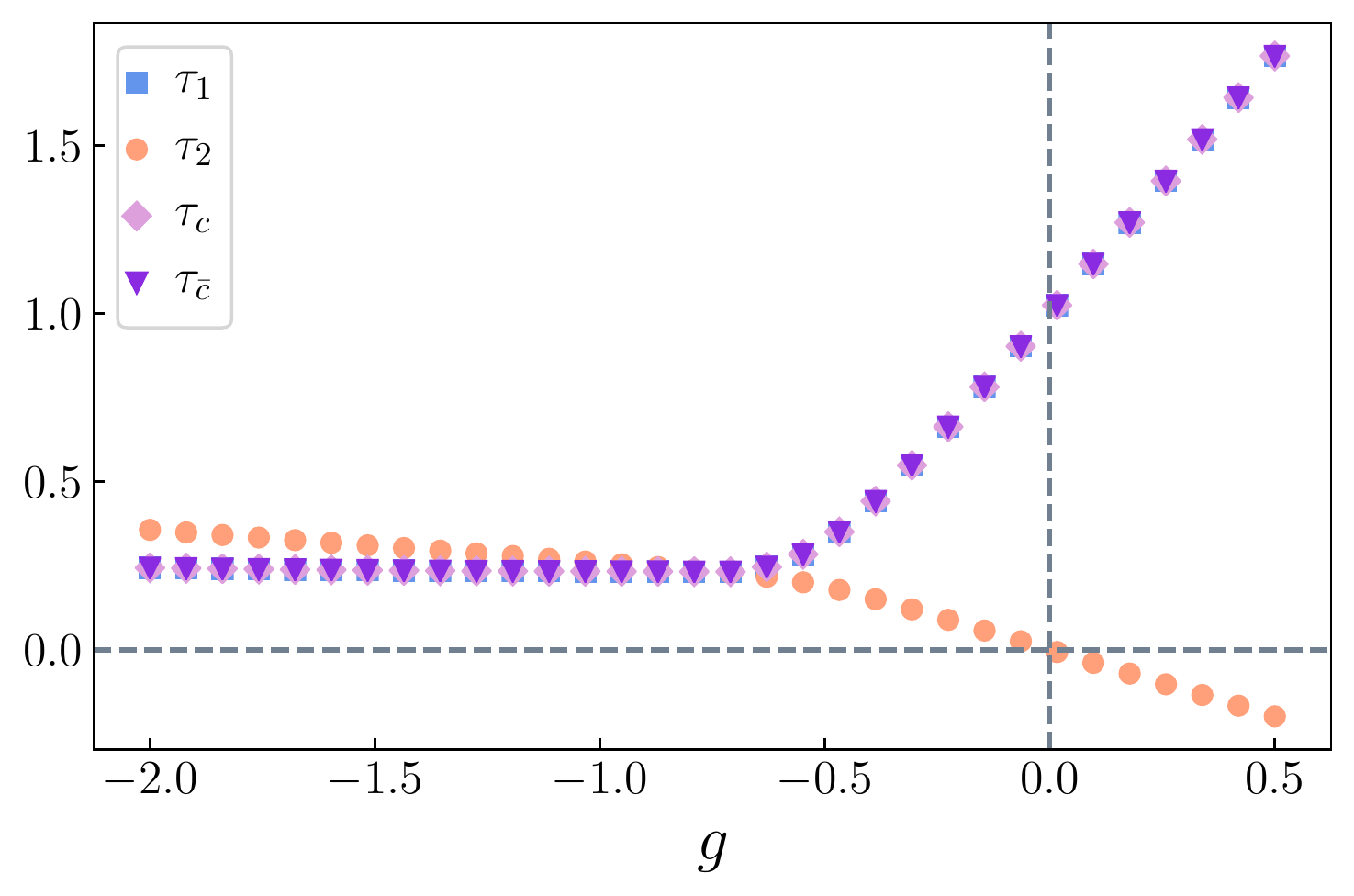}
    \caption{The values of $\{\tau_j\}$ determined self-consistently using Eq.\,\eqref{eq:taus}. In the non-interacting limit $\tau_1 = \tau_{\bar{c}} = \tau_{c} = t$, as expected, and any finite $g$ renormalizes the nearest neighbor tunneling amplitudes, while introducing next nearest neighbor hopping. The spectrum of \eqref{eq:HMF} is gapped for all $g$, except at the mean field critical point $\gMF \approx -0.73$ where curves intersect.}
    \label{fig:tri_SC_MF}
  \end{figure}

  It is interesting to note that the first relation in \eqref{eq:taus} implies that $\tau_c = \tau_{\bar{c}}$ for any $g$ and, thereby, precludes an ordered phase. To understand this result, consider the following argument. The square lattice Majorana-Hubbard model involves only one kind of plaquette and pairing MZMs along a given direction minimizes the energy of one-half of the plaquettes. At strong enough interactions, such a state is favored by the system as a whole \cite{affleck2017majorana}. In the present case, we have three kinds of plaquettes $\mP_\nu$. If the MZMs were to dimerize along $\bm{c}$, say, then the energy of one-half of $\mP_1$ would be minimized. Nevertheless, as these plaquettes constitute only one-sixth of the total number of interaction terms, such a configuration is energetically unfavorable. In this manner, the geometric frustration inherent to the triangular lattice distinguishes it from a square lattice of Majoranas.

\begin{figure}
    \centering
    \includegraphics[width=8.5cm]{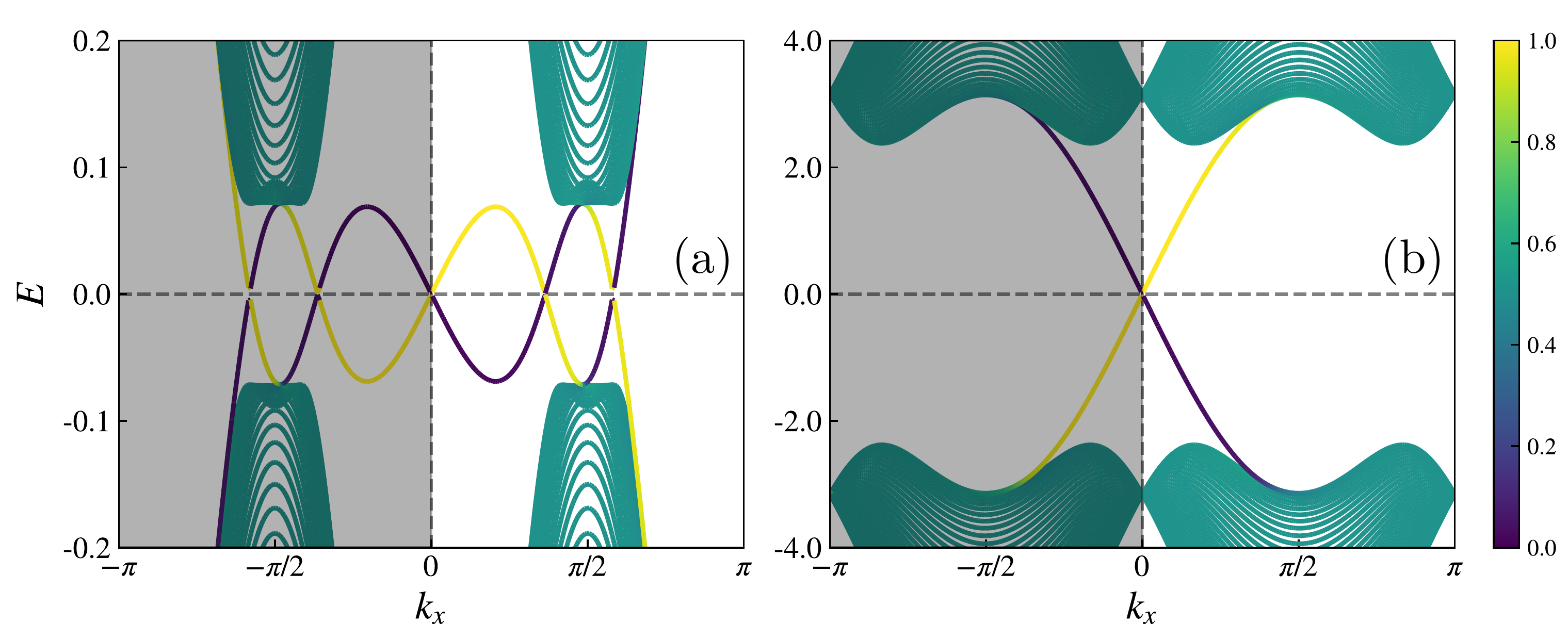}
    \caption{The spectra of $\mHMF$ on a strip for self-consistent $\{\tau_j\}$ when (a) $g=-1.0$ and (b) $g=0.5$. As before, the color scale denotes the normalized expectation value of the $\hat{y}$ position operator and only one-half of the Brillouin zone is physical. The edge states show that the transition at $\gMF$ separates phases with $\mathcal{C} = 3$ and $\mathcal{C} = 1$.}
    \label{fig:tri_MF_edges}
  \end{figure}

  The self-consistent $\{\tau_j\}$, shown in Fig.\,\ref{fig:tri_SC_MF}, further indicate that all the effective nearest neighbor hopping amplitudes vary identically and $\tau_2$ is nonzero at any finite coupling. For $g > \gMF \approx -0.73$, the mean field spectrum is gapped. In the absence of interactions, $\mHMF$ reduces to $\mH$ because $\tau_c = \tau_{\bar{c}} = \tau_1 = t$ and $\tau_2 = 0$. Based on this, we deduce that for $g>\gMF$, $\mHMF$ is in the same topological phase as $\mH$. At $\gMF$, all $\tau_j$ coincide and the dispersion exhibits two quadratic band crossings, as noted in Appendix \ref{sec:H_MF}. In general, the Berry flux at a quadratic touching is either $0$ or $\pm 2\pi$. The spectra in Fig.\,\ref{fig:tri_MF_edges} show that the latter holds true here and $\mathcal{C}=3$ for $g<\gMF$. In other words, $\gMF$ marks a topological phase transition.

\section{Numerical phase diagram} \label{sec:4leg_lad}

  Beyond mean field theory, analytical techniques to study the model are scarce. In order to address the phase diagram while fully accounting for the quantum correlations, we rely on DMRG. The simplest variation of the 2D model that includes all three kinds of plaquettes while being amenable to numerics is a ladder with four legs, which is equivalent to $N_y = 2$ in our notation.

  To implement the Hamiltonian, we map the Majorana degrees of freedom to spinless Dirac fermions, which provide a formally equivalent yet more convenient representation (see Appendix \ref{sec:H_cf}). Since two Majorana modes compose one Dirac fermion, in the new basis one obtains a two-leg ladder with $N = 2N_x$ fermions. The model with open boundaries, as discussed previously, exhibits edge states that interfere with the determination of the bulk gaps. To circumvent this, we focus on tori with periodicity along $y$ and anti-periodic boundary conditions along $x$ -- this choice is found to be helpful for the purpose of converging on the excited states. As a check, the DMRG code has been benchmarked against exact diagonalization for small system sizes.

\begin{figure}
    \centering
    \includegraphics[width=8cm]{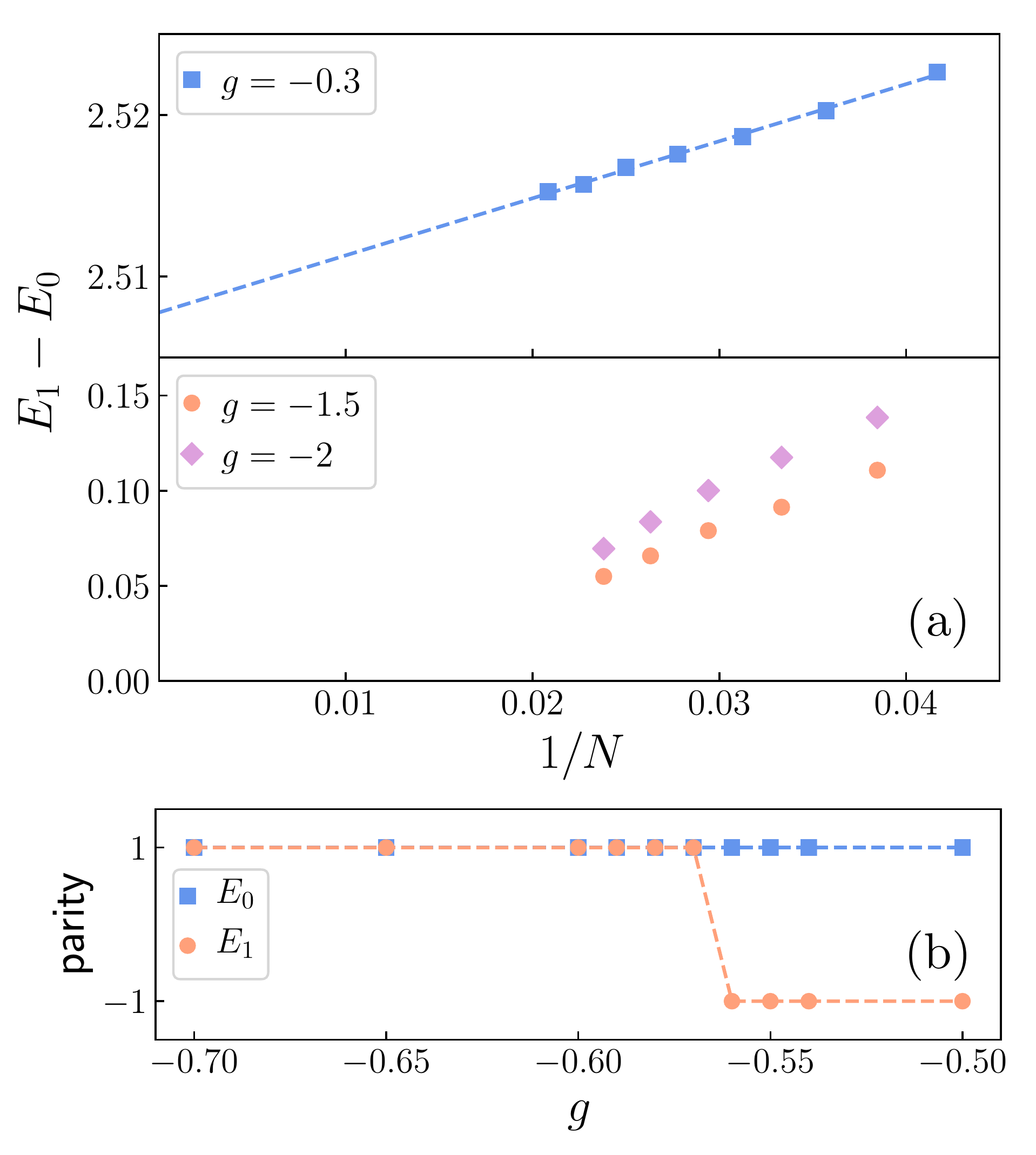}
    \caption{(a) The energy gap as a function of inverse system size for different values of the coupling $g$. The dashed line is a linear fit to the data. A bond dimension of up to $1100$ states in the DMRG sweeps ensures that the truncation errors are less than $10^{-7}$. The largest system size considered is $N=48$, which corresponds to $96$ Majorana modes. (b) Fermionic parity of the two lowest states in the spectrum as a function of interaction strength.}
    \label{fig:gaps}
  \end{figure}

  We begin by studying the gap to the first excited state as a function of system size. On the basis of the mean field analysis one would anticipate the spectrum to be gapped for different values of coupling strength, except possibly at the transition. Unexpectedly, signatures of a gapless phase emerge for a range of attractive interactions, as seen in Fig.\,\ref{fig:gaps}(a). An interesting feature is that the gaps are system size dependent: in the gapless phase, when $N_x$ is even, for instance, the first excited state is exactly degenerate with the ground state, so for the sake of clarity only gaps corresponding to odd $N_x$ are shown. While larger systems would be ideal in ascertaining the vanishing gap, the cons of imposing periodic boundaries, in conjunction with the fact that fermionic parity is the only symmetry at our disposal, limit the accessible system sizes.

  In addition to the gap, another quantity that distinguishes the two phases is parity of the first excited state. As shown in Fig.\,\ref{fig:gaps}(b), it switches from odd to even as $g$ is reduced. Treating this as a criterion, the transition can be identified at $g_c \approx -0.56$. Moreover, as discussed in Appendix \ref{sec:ener_deriv}, behavior of the ground state energy, and its derivatives, with respect to the coupling suggests that the transition is of second order.


  \subsection{Central charge and transition}

    \begin{figure}
      \centering
      \includegraphics[width=8cm]{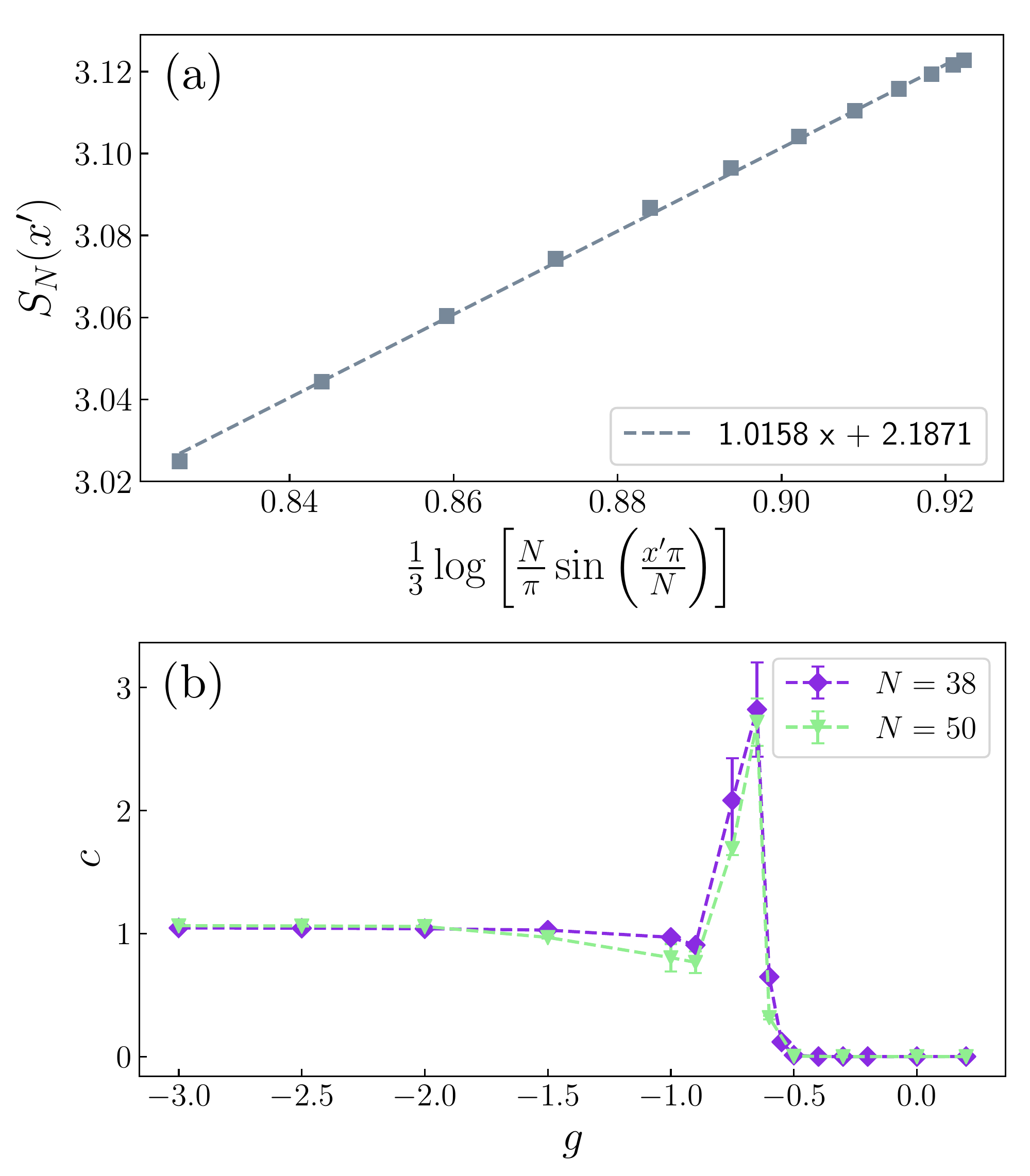}
      \caption{(a) Entanglement entropy, which has been averaged over neighboring bonds (see text), in a periodic system of length $N = 50$ at $g = -3$. The conformal distance is plotted on the horizontal axis and the dashed line is a linear fit whose slope corresponds to $c$. (b) Central charge as a function of the coupling. We find that $c$ remains unity for values as low as $g=-20$.}
      \label{fig:cnt_chg}
    \end{figure}

    Since the model is quasi one-dimensional and gapless, we might expect low energy behavior in the critical phase to be represented by a conformal field theory (CFT). An important quantity that characterizes a CFT is its central charge $c$, which can be thought of as a measure of the gapless degrees of freedom. If a periodic system of size $N$ is described by a CFT with central charge $c$, then the entanglement entropy of a subregion of size $x$ in the ground state is predicted to scale as \cite{calabrese2009entanglement}
    \begin{align}
      S_N(x) = \frac{c}{3} \log{\left[\frac{N}{\pi} \sin\left(\frac{x \pi}{N}\right) \right]} + S_0,
    \end{align}
    where $S_0$ is a non-universal constant. The two-site unit cell results in an oscillatory $S_N(x)$. Averaging the entropy across neighboring bonds ($x$ and $x+1$) and assigning it to the middle ($x' = x+1/2$) eliminates the oscillatory sub-leading terms and aids the determination of $c$ \footnote{Alternatively, to get around the oscillations, one could simply fit the entropy on even (or odd) bonds. Indeed, we find that this gives the same value of central charge.}. Following this prescription, we find that the gapless phase belongs to the moduli space of $c=1$ conformal theories. While the value of central charge in the two extended phases is unambiguous, its behavior in the vicinity of the transition is more difficult to establish. These findings are summarized in Fig.\,\ref{fig:cnt_chg}.

\begin{figure}
      \centering
      \includegraphics[width=8cm]{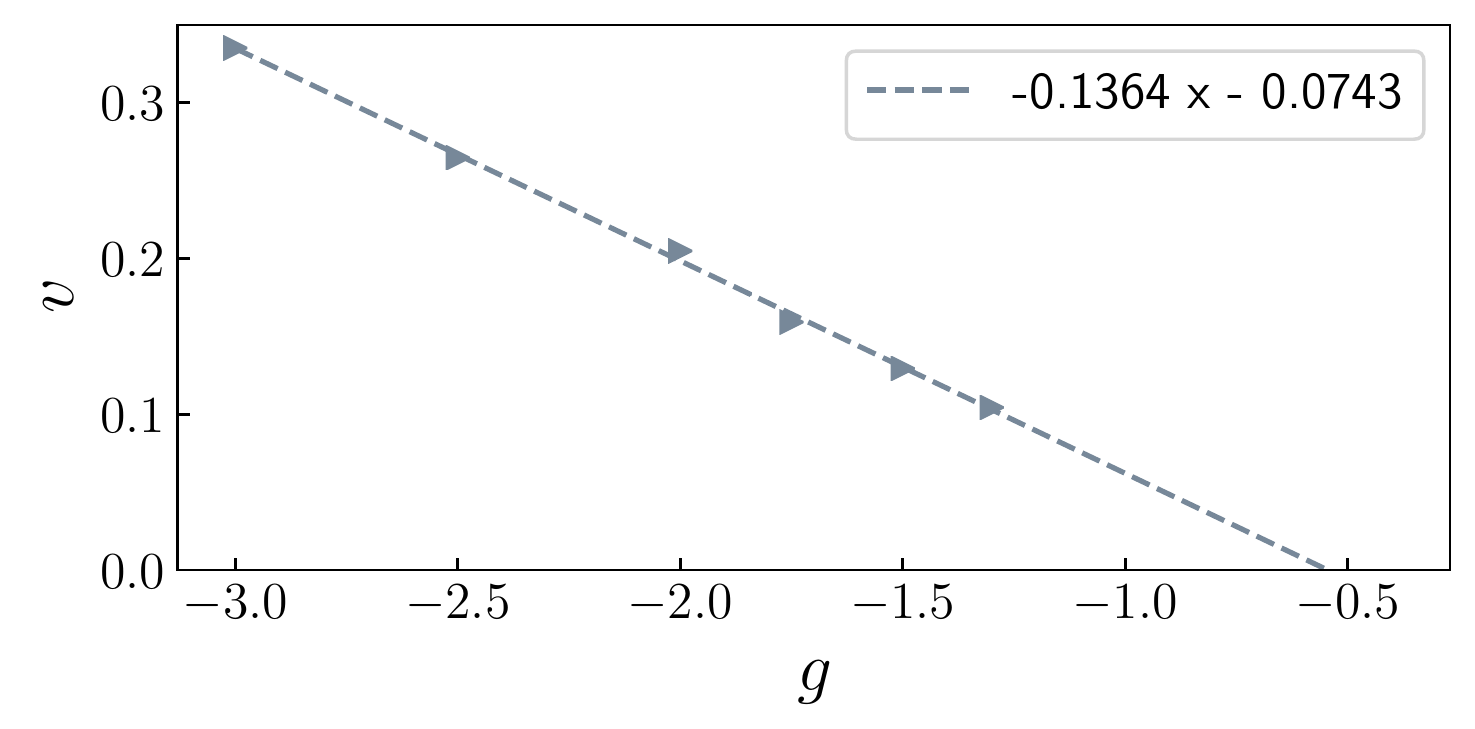}
      \caption{The velocities, obtained using Eq.\,\eqref{eq:vel}, for different values of coupling in the vicinity of the transition. The dashed line is a linear fit to the data.}
      \label{fig:vel_vs_g}
    \end{figure}

    To shed some light on the nature of the transition, we approach it from within the critical phase. The velocity of excitations, which describes the linearized CFT spectrum at Fermi energy, can be estimated from a finite-size scaling of the ground state energy. Assuming that all excitations propagate with the same velocity $v$, the energy density of a periodic system is given by \cite{blote1986conformal}
    \begin{align}
      \frac{E_0}{N} = e_{\infty} - \frac{\pi c v}{6 N^2} + \dots,
      \label{eq:vel}
    \end{align}
    where $e_{\infty}$ is the ground state energy per site in the thermodynamic limit and the ellipsis denotes finite size corrections. The numerically determined velocities are shown in Fig.\,\ref{fig:vel_vs_g}. Observe that $v$ vanishes as one approaches the phase transition and a linear fit gives $g_c \approx -0.54$, which collates well with the critical value signaled by the change in parity of the first excited state. A reliable extraction of $v$ closer to the critical point is complicated by the fact that the above procedure relies on a precise knowledge of the central charge. More sophisticated methods would be necessary to further characterize the phase transition.

\begin{figure}
    \centering
    \includegraphics[width=7cm]{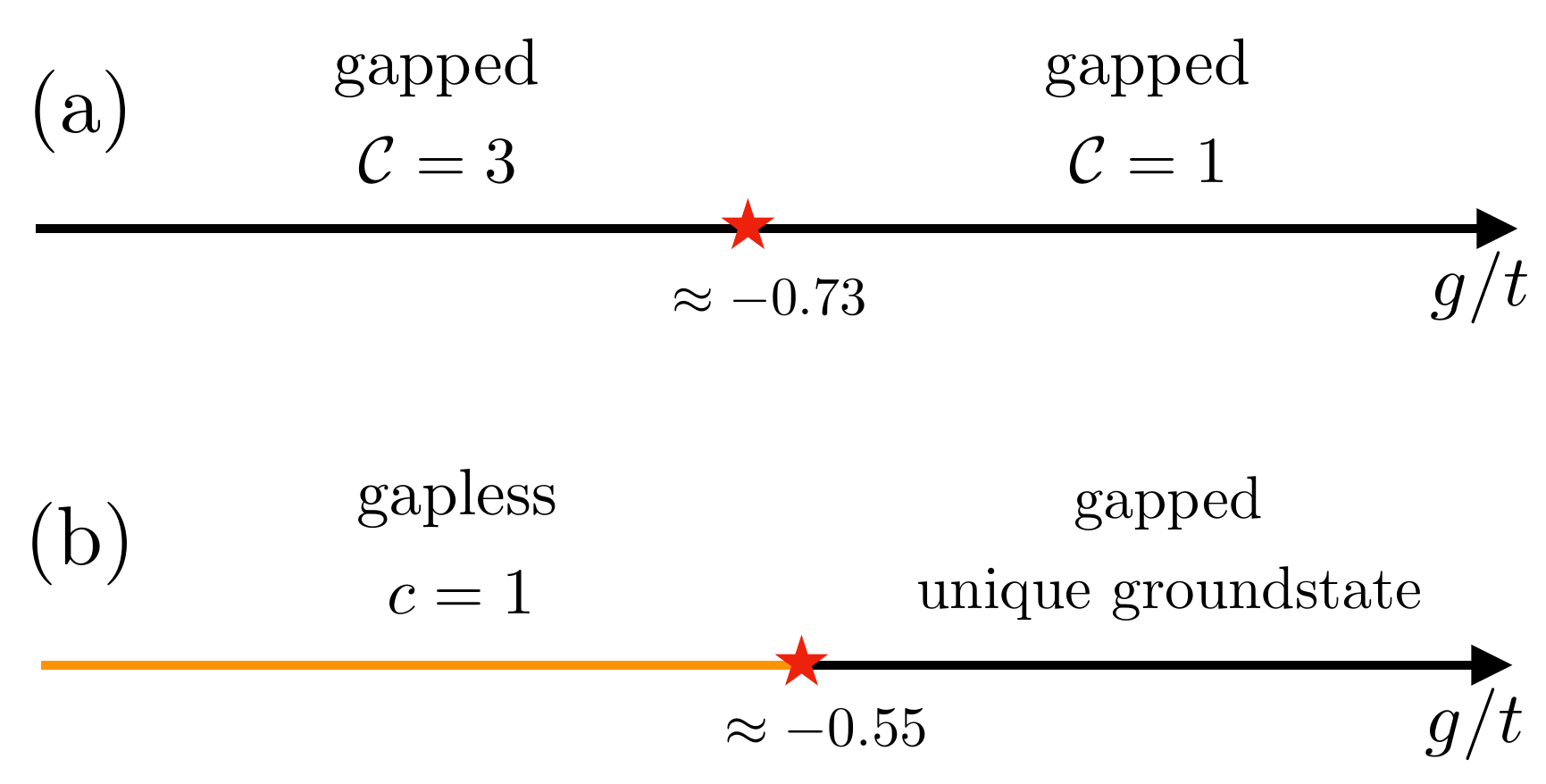}
    \caption{Comparison of (a) mean field and (b) DMRG phase diagrams of the model. The stars denote the critical values at which phase transitions occur.}
    \label{fig:phase_diag}
  \end{figure}

    Collectively, the results presented in this section suggest that when the interactions are attractive in nature, the physics of a thin torus deviates from the mean field predictions. In particular, an extended critical phase replaces a gapped topological phase. The mean field and DMRG phase diagrams are sketched in Fig.\,\ref{fig:phase_diag}.


\section{Conclusions}
  There has been a growing body of evidence for MZMs in experiments \cite{Xu_2015, Zhang_2018}. Recent scanning tunneling microscopy studies of iron-based superconductors have observed distinct zero-bias peaks at the cores of vortex defects \cite{wang2018evidence, Liu_2018, machida2019zero}. In the light of these developments, we have explored a Hubbard like tight-binding model aimed at providing a low energy description of zero modes in the experimentally pertinent triangular vortex lattice.

  In the absence of interactions, the model is a gapped Majorana Chern insulator. A self-consistent mean field analysis suggests that this phase persists for repulsive interactions and there are no signatures of spontaneous symmetry breaking. Strong enough attractive interactions, on the other hand, bring about a topological transition into a phase with a higher Chern number. Numerical simulation of tori agrees with the mean field picture for the repulsive regime. When the coupling is tuned to a critical value, however, a gapless phase emerges. It would be reasonable to suspect that some of the details are artifacts of working with a small linear dimension along one direction. In this regard, a comparative study of tori with larger width is an interesting avenue for future studies.

  \emph{Acknowledgements.} We thank Marcel Franz, \'Etienne Lantagne-Hurtubise and Chengshu Li for helpful discussions and comments on the manuscript. This work was supported by the Max Planck-UBC-UTokyo Centre for Quantum Materials and the Canada First Research Excellence Fund, Quantum Materials and Future Technologies Program. T.T. and I.A. acknowledge support from NSERC through Discovery Grant No. 04033-2016. Part of the numerical work described here was performed using the ITensor library \cite{ITensor}.


\appendix

\section{Hamiltonian and complex fermion representation} \label{sec:H_cf}

  Here we introduce a handy notation to denote the zero mode operators. The two sites belonging to a unit cell located at the position vector $m \bd_1 + n \bd_2$ are labeled by $\gamma^{r}_{m,n}$ and $\gamma^{b}_{m,n}$, corresponding to the red and blue sub-lattices respectively. Here, $\bd_1$ and $\bd_2$ are the Bravais lattice vectors as before. Periodicity along $x$ and $y$ corresponds to $\gamma^{r}_{N_x+m, N_y+n} = \gamma^{r}_{m,n}$. Similar relation holds for the other sub-lattice.

  As per the gauge depicted in Fig.\,\ref{fig:lattice1}(a), the non-interacting Hamiltonian reads
  \begin{align}
    \mathcal{H}_0 = it \sum_{m,n}
      & [\gamma^{r}_{m,n} (\gamma^{r}_{m+1,n} + \gamma^{b}_{m,n} - \gamma^{b}_{m-1,n})  \\
      & - \gamma^{b}_{m,n} (\gamma^{b}_{m+1,n} + \gamma^{r}_{m+1,n+1} + \gamma^{r}_{m,n+1})]. \nonumber
  \end{align}

  And the three kinds of plaquette interactions that compose $\mHI$ are given by
  \begin{align}
    \mathcal{P}_1 = \sum_{m,n}
        [& \gamma^{r}_{m,n} \gamma^{r}_{m-1,n} \gamma^{b}_{m-2,n} \gamma^{b}_{m-1,n} \nonumber \\
        & + \gamma^{b}_{m,n} \gamma^{b}_{m-1,n} \gamma^{r}_{m-1,n+1} \gamma^{r}_{m,n+1}] \\
    \mathcal{P}_2 = \sum_{m,n}
       [& \gamma^{r}_{m,n} \gamma^{b}_{m-1,n-1} \gamma^{r}_{m,n-1} \gamma^{b}_{m,n-1} \nonumber \\
        & + \gamma^{b}_{m,n} \gamma^{r}_{m,n} \gamma^{b}_{m,n-1} \gamma^{r}_{m+1,n}] \\
    \mathcal{P}_3 = \sum_{m,n}
       [& \gamma^{r}_{m,n} \gamma^{r}_{m+1,n} \gamma^{b}_{m+1,n} \gamma^{b}_{m,n} \nonumber \\
       & + \gamma^{b}_{m,n} \gamma^{b}_{m+1,n} \gamma^{r}_{m+2,n+1} \gamma^{r}_{m+1,n+1}].
  \end{align}

  When the two zero modes in a unit cell are combined into a complex fermion, we have
  \begin{align}
    & \gamma^{r}_{m,n} = c_{m,n}^\dagger + c_{m,n} \nonumber \\
    & \gamma^{b}_{m,n} = i (c_{m,n}^\dagger - c_{m,n}).
  \end{align}
  The full Hamiltonian may now be expressed in this basis.


\section{Symmetries and gauge transformations} \label{sec:app_symm}

  In terms of the operators $\gamma^{r}_{m,n}$ and $\gamma^{b}_{m,n}$, the symmetries of the model are as follows:

  \emph{Translations.}
    The action of $\mathcal{T}_\mathbf{a}$ corresponds to translation by a unit cell along the $x$ axis and it is clearly a symmetry. Translation by a site along $\mathbf{c}$ also leaves the Hamiltonian invariant, provided that it is accompanied by the gauge transformation
    \begin{align}
      \gamma^{r}_{m,n} & \to (-1)^{m+n} \gamma^{b}_{m,n} \nonumber \\
      \gamma^{b}_{m,n} & \to (-1)^{m+n+1} \gamma^{r}_{m+1, n+1}.
    \end{align}


  \emph{Reflections and time reversal.}
    Assuming that the $x$ axis passes through the sites $\gamma^{r}_{m,0}$, the product of reflection about $x$ and time reversal, $\mathcal{R}_{x} \Theta$, is given by
    \begin{align}
      \gamma^{r}_{m,n} & \to (-1)^{m} \gamma^{r}_{m, -n} \nonumber \\
      \gamma^{b}_{m,n} & \to (-1)^{m+1} \gamma^{b}_{m, -n-1}  \nonumber \\
      i & \to -i.
    \end{align}

    With the convention that the $\gamma^{r}_{0,n}$ sites lie on the $y$ axis, the combination $\mathcal{R}_{y} \Theta$ corresponds to
    \begin{align}
      \gamma^{r}_{m,n} & \to (-1)^{n} \gamma^{r}_{-m, n} \nonumber \\
      \gamma^{b}_{m,n} & \to (-1)^{n} \gamma^{b}_{-m-1, n} \nonumber \\
      i & \to -i.
    \end{align}


  \emph{Rotation by $\pi/3$.}
    The six-fold rotation symmetry interchanges the two sub-lattices in a manner that is dependent on the position of the sites with respect to the rotation center. For clarity, we switch to a basis that is natural to the triangular geometry and label the sites with the vectors $\br = m \bm{p} + n \bm{q}$, where $\mathbf{p} = (1,0)$ and $\mathbf{q} = ({1 / 2}, {\sqrt{3} / 2})$. The Hamiltonian \eqref{eq:H0} in this notation reads
    \begin{multline}
      \mH = it \sum_{m,n}  [(-1)^n \gamma_{m,n} \gamma_{m+1,n}
                   + (-1)^n \gamma_{m,n} \gamma_{m,n+1} \\
                   - \gamma_{m,n} \gamma_{m-1,n+1}].
      \label{eq:H_rot_sym}
    \end{multline}
    A clockwise rotation by $\pi/3$ corresponds to the transformation $\gamma_{m,n} \to s_{m,n} ~ \gamma_{m+n, -m}$, with
    \begin{equation}
      s_{m,n} =
      \begin{cases}
        (-1)^m (-1)^{\frac{n-1}{2}}, & \text{if $n$ odd} \\
        (-1)^{\frac{n}{2}}, & \text{if $n$ even}.
      \end{cases}
      \label{eq:rot}
    \end{equation}
    It is a simple exercise to check that \eqref{eq:H_rot_sym} is invariant under this.


\section{Mean field self-consistency equations} \label{sec:app_mft}

  Assuming translational invariance and employing Wick's theorem, the energy density of the full Hamiltonian in the state $\ket{\psiMF}$ can be evaluated as
  \begin{align}
    {\Braket{\mathcal{H}} \over N} = t & (4\Delta_1 + \Delta_c + \Delta_{\bar{c}})
      + g [\Delta_2 (4 \Delta_1 + \Delta_{c} + \Delta_{\bar{c}})]  \nonumber \\
      & - g [8\Delta_1^2 + \Delta_{c}^2 + \Delta_{\bar{c}}^2 + 2 \Delta_{c} \Delta_{\bar{c}}].
  \end{align}
  where $N=N_x N_y$ is the system size. The goal is to find $\tau_i$ ($i \in \{c, \bar{c}, 1, 2\}$) that satisfy $\partial \Braket{\mathcal{H}}/\partial \tau_{i} = 0$, that is,
  \begin{align}
    &
    [t - g (2 \Delta_c + 2 \Delta_{\bar{c}} - \Delta_2)]
    \left( \frac{\partial \Delta_c}{\partial \tau_i} + \frac{\partial \Delta_{\bar{c}}}{\partial \tau_i} \right)
    \nonumber \\
    + & 4[t - g (4 \Delta_1 - \Delta_2)] \frac{\partial \Delta_1}{\partial \tau_i}
    \nonumber \\
    + & g (4 \Delta_1 + \Delta_c + \Delta_{\bar{c}}) \frac{\partial \Delta_2}{\partial \tau_i}
    = 0.
    \label{eq:H_delta_cond}
  \end{align}
  In order to connect this with the mean field Hamiltonian, note that the definition of $\mHMF$ motivates an alternate expression for $\Delta_j$. Namely,
  \begin{align}
    \Delta_{j} = \frac{1}{\rho_j N} \Braket{\frac{\partial \mHMF}{\partial \tau_{j}}}
    = \frac{1}{\rho_j N} \frac{\partial E_{\text{MF}}}{\partial \tau_{j}},
    \label{eq:del_der}
  \end{align}
  where we have defined the bond dependent constants $\rho_c = \rho_{\bar{c}}=1$, $\rho_1 = 4$ and $\rho_2 = 6$. Observe that $\Delta_j$ have been related to $E_{\text{MF}}$ via the Hellmann-Feynman theorem, which associates the expectation value of a derivative of an operator with the derivative of its expectation value. And from the definition \eqref{eq:HMF} we know that
  \begin{align}
    {E_{\text{MF}} \over N} = {\Braket{\mHMF} \over N}
    = \tau_c \Delta_c + \tau_{\bar{c}} \Delta_{\bar{c}} + 4 \tau_1 \Delta_1 + 6 \tau_2 \Delta_2.
  \end{align}
  Simplifying the expression for $\partial E_{\text{MF}}/ \partial \tau_{i}$ using \eqref{eq:del_der} leads to
  \begin{align}
      \tau_c \frac{\partial \Delta_c}{\partial \tau_i}
    + \tau_{\bar{c}} \frac{\partial \Delta_{\bar{c}}}{\partial \tau_i}
    + 4 \tau_1 \frac{\partial \Delta_1}{\partial \tau_i}
    + 6 \tau_2 \frac{\partial \Delta_2}{\partial \tau_i} = 0.
    \label{eq:HMF_delta_cond}
  \end{align}
  Finally, by comparing \eqref{eq:H_delta_cond} with the above relation \eqref{eq:HMF_delta_cond}, one obtains the expressions in Eq.\,\eqref{eq:taus}.


\section{Mean field spectrum} \label{sec:H_MF}

  In momentum space, the mean field Hamiltonian can be written as
  \begin{align}
    \mHMF = \sum_{\bk}{}^{'}
    \Psi^{\dagger}_{\bk}
    \begin{pmatrix}
        D_1(\bk) & D_2(\bk) \\[6pt]
        D_2(\bk)^* & -D_1(\bk)
    \end{pmatrix}
    \Psi_{\bk}
    \label{eq:H_MF_k}
  \end{align}
  where $\Psi_{\bk} = (\gamma^{r}_{\bk}, \gamma^{b}_{\bk})^{\text{T}}$, the sum is restricted to half of the Brillouin zone, and
  \begin{align}
    D_1(\bk) & = -4 [\tau_1 \sin (\bk \cdot \bd_1) + \tau_2 \sin (\bk \cdot \bd_2)]  \\
    D_2(\bk) & = 2i[\tau_c + \tau_{\bar{c}} e^{-i \bk \cdot (\bd_1 + \bd_2)}
                  + \tau_1 (- e^{-i \bk \cdot \bd_1} + e^{-i \bk \cdot \bd_2}) \nonumber \\
                  + \tau_2 & (e^{i \bk \cdot (\bd_1 - \bd_2)} - e^{i \bk \cdot \bd_1}
                            + e^{- 2i \bk \cdot \bd_1} + e^{-i \bk \cdot (2\bd_1 + \bd_2)})].
                  \nonumber
  \end{align}
  The spectrum, $\pm \sqrt{D_1(\bk)^2 + |D_2(\bk)|^2}$, displays a finite gap, except when $\tau_1 = \tau_c = \tau_{\bar{c}} = \pm \tau_2$. At the mean field critical point $\gMF$, $\tau_j = \tau$ for all $j$ and, consequently, the dispersion simplifies to
  \begin{align}
    E_{\rm{MF}}^{\pm}&(\bk) = \pm 2 \sqrt{2} \tau
    \biggl[
    6 + 3 \cos\left(2 k_x\right) - \cos\left(2 \sqrt{3} k_y\right)
    \nonumber \\
    & + 4 \sin\left(k_x\right) \sin\left(\sqrt{3} k_y\right)
    \left(2 + \cos\left(2 k_x\right) \right)
    \biggr]^{1/2}.
  \end{align}
  Clearly, the energy vanishes at $\bk = \pm(\pi/2, -\pi/2\sqrt{3})$. An expansion in small momenta in the vicinity of these nodes shows that the dispersion is in fact quadratic.


\begin{figure}[h!]
  \centering
  \includegraphics[width=8.5cm]{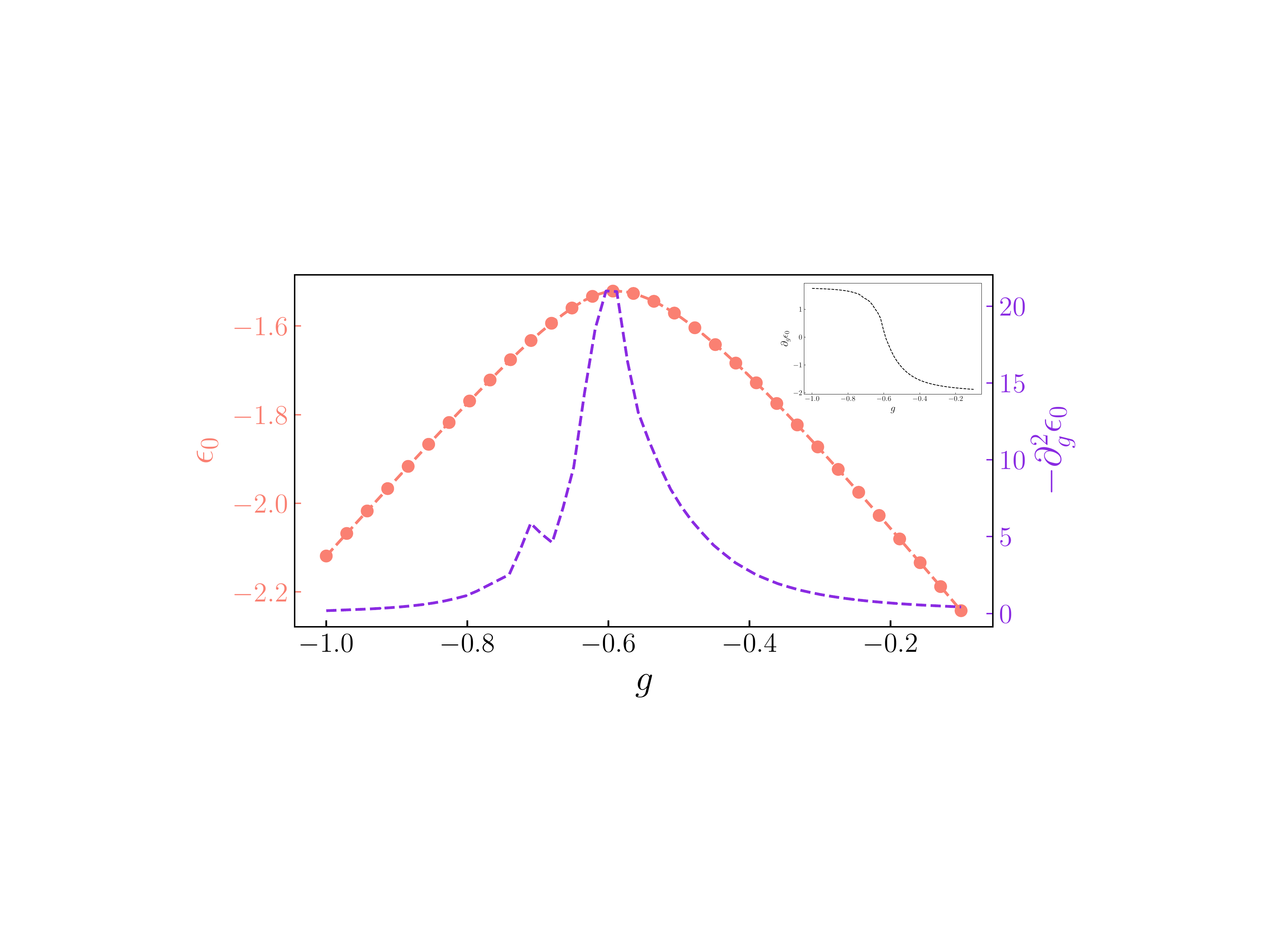}
  \caption{The ground state energy density (left axis) and its second derivative (right axis) with respect to the coupling, obtained with $N=30$. Inset: profile of the first derivative in the same range of $g$.}
  \label{fig:ener_deriv}
\end{figure}

\section{Order of the phase transition} \label{sec:ener_deriv}

  While the first excited state shows a change in parity at $g_c$, signatures of the phase transition can also be found in the lowest energy state. The ground state energy density $\epsilon_0 = E_0/N$ is a thermodynamic quantity and a discontinuity in its derivatives indicates the order of the transition. We find that the second derivative of $\epsilon_0$ shows indications of discontinuity very close to the critical point ($g \approx -0.58$), as shown in Fig.\,\ref{fig:ener_deriv}.


\bibliography{mhtl}

\end{document}